\documentclass[prc,twocolumn,showpacs,preprintnumbers,amsmath]{revtex4}
\usepackage{dcolumn}
\usepackage{bm}
\usepackage{graphicx}
\begin{document}
\title{Phase transition and properties of compact star}
\author{B.K. Sharma} 
\author{P.K. Panda} 
\author{S.K. Patra} 
\affiliation{Institute of Physics, Sachivalaya Marg, Bhubaneswar -
751 005, India.}
\begin{abstract}
We investigate the phase transition to a deconfined phase and
the consequences in the formation of neutron stars. We use
the recently proposed effective field theory motivated
relativistic mean field theory for
hadron and the MIT Bag model and color-flavor locked (CFL) phase 
for the quark matter in
order to get the appropriate equation of state.
The properties of star are then calculated. The differences
between unpaired and CFL quark matter are
discussed.
\end{abstract}
\maketitle

\section{Introduction}
One of the fundamental problem of nuclear physics is to understand
the behavior of nuclear matter at extreme conditions. The study of
neutron star provide an important information in this regard. Neutron
star are extremely dense objects. They formed after the
gravitational core collapse of a massive star \cite{Gl97}. The
environment in central region resembles the early universe, except
that the temperature (T) is lower. Due to the lower temperature and high
density, neutron stars are presumably unique astrophysical
laboratories for a broad range of physical phenomena
\cite{Gl97,Sha83}. The composition and other properties of neutron
stars are depend upon the appropriate equation of state (EOS) that
describe its crust and interior region \cite{Da02}. The crust part
of neutron star, where the density is comparable to saturation
density ($\rho_0$) of symmetric nuclear matter, is adequately
described by hadronic matter. But the interior region where density
$\rho$ is of the order of 5 to 10 times of $\rho_0$, a phase
transition occurs from hadronic to quark matter. But this is not
well understood, whether the interface region, it is a quark and/or
mixed matter. It is reasonable to assume that the central part is
dense enough, so that it can be treated as a pure quark phase.
However, the boundary part between the quark and hadronic matter
should be an admixture of quark and baryonic degrees of freedom. In
the present study, we are interested in building the EOS for mixed
matter of quark and hadron phases. We employ an effective field
theory motivated (E-RMF) Lagrangian approach including hyperons in
order to describe the hadron phase. For the quark phase we have
chosen to use both unpaired quark matter (UQM) described by the MIT
bag model \cite{bag,freedman,farhi,kapusta} and paired quarks
described by the color-flavor locked (CFL) phase. Recently many
authors \cite{shh,ar,bo,srp,arrw,kw,pkp} have discussed the possibility
that the quark matter is in a color superconducting phase, in which
quarks near the Fermi surface are paired, forming Cooper pairs which
condense and break the color gauge symmetry \cite{mga}. At
sufficiently high density the favored phase is called CFL, in which
quarks of all three colors and all three flavors are allowed to
pair.

The paper is organized as follows: in section II the E-RMF model for
hadronic phase, MIT bag model and CFL phase for quark matter and
mixed phase are described. Here, we are trying to build an equation
of state for mixed matter of hadron and quark phases. We use an
E-RMF model including hyperons with incompressibility $K= 300$ MeV
and $M^*/M= 0.7$ of nuclear matter for hadron phase. We use Gibbs
criteria and chemical equilibrium conditions, to built a mixed phase
EOS, and then we calculate and discuss the properties of star at
T=0. The calculated results are discussed in section III. In the
last section, the conclusions are drawn.

\section{Formalism}
\subsection{Hadron matter}
In principle, one should use quantum chromodyanmics (QCD), the
fundamental theory of strong interaction, for the complete
description of EOS. But it cannot be use to describe hadronic matter
due to its non-perturbative properties. A major breakthrough
occurred when the concept of effective field theory (EFT) was
introduced and applied to low energy QCD \cite{We79}. The EFT for
strong interaction at low energy is known as quantum hadrodynamics
(QHD) \cite{Wa74,Se86,Re89,Se92,Se97}. The mean field treatment of
QHD has been used extensively in order to describe properties of
nuclear matter \cite{Wa74,Aru04} and finite nuclei
\cite{Ho81,Bo77,Ga90,Ri97}. The degrees of freedom in this theory
are nucleons interacting through the exchange of iso-scalar scalar
$\sigma$, iso-scalar vector $\omega$, iso-vector-vector $\rho$ and
the pseudoscalar $\pi$ mesons. The nucleons are considered as Dirac
particle moving in classical meson fields. The contribution of $\pi$
meson is zero at mean field level, due to pseudo-spin nature. The
chiral effective Lagrangian (E-RMF) proposed by Furnstahl, Serot and Tang
\cite{Fu96,Fu97,Mu96} is the extension of the standard relativistic
mean field (RMF) theory with the addition of non-linear
scalar-vector and vector-vector self interaction. This Lagrangian
includes all the non-renormalizable couplings consistent with the
underlying symmetries of QCD. Applying the naive dimensional
analysis \cite{Ge84,Ge93} and the concept of naturalness one can
expand the nonlinear Lagrangian and organize it in increasing powers
of the fields and their derivatives and truncated at given level of
accuracy \cite{Fu97b,Fu00,Se04}. In practice, to get a reasonable
result, one needs the Lagrangian up to 4th order of interaction. In
the interior of neutron star, where the density is very high, other
hadronic states are produced \cite{Gl97,Gl85,Be01,Sch99}. Thus, the
considered model involves the full octet of baryons interacting
through mesons. The truncated Lagrangian which includes the terms up
to the fourth order is given by

\begin{widetext}
\begin{eqnarray}
{\cal L} & = & \overline{\Psi}_{B}\left ( i\gamma^\mu
D_{\mu} - m_{B} + g_{\sigma B}{\sigma}\right)
{\Psi}_{B}
+\frac{1}{2}{\partial_{\mu}}{\sigma}{\partial^{\mu}}{\sigma}
-m_{\sigma}^2{\sigma^2}\left(\frac{1}{2}
+\frac{\kappa_3}{3 !}
\frac{g_{\sigma B}\sigma}{m_{B}}+\frac{\kappa_4}{4 !}
\frac{g^2_{\sigma B}\sigma^2}{m_{B}^2}\right)
-\frac{1}{4} {\Omega_{\mu\nu}}{\Omega^{\mu\nu}}
\nonumber \\
& & +\frac{1}{2}\left (1 +
{\eta_1}\frac{g_{\sigma B}\sigma}{m_{B}}
+\frac{\eta_2}{2}\frac{g^2_{\sigma B}\sigma^2}{m_{B}^2}\right)
m_{\omega}^2{\omega_{\mu}}
{\omega^{\mu}}
- \frac{1}{4}{R^a_{\mu\nu}}{R^{a\mu\nu}}
+\left(1 + \eta_{\rho}\frac{g_{\sigma B}\sigma}{m_{B}}\right)
\frac{1}{2}m_{\rho}^2{\rho^a_{\mu}}{\rho^{a\mu}}+\frac{1}{4 !}{\zeta_{0}}
g^2_{\omega B}\left({\omega_{\mu}}{\omega^{\mu}}\right)^2
\end{eqnarray}
\end{widetext}
The subscript
$B=n,p,\Lambda,\Sigma$ and $\Xi$, denotes for baryons. The terms in
eqn. (1) with the subscript $"B"$ should be interpreted as sum over
the states of all baryonic octets. The covariant derivative ${D_{\mu}}$
is defined as
\begin{eqnarray}
{D_{\mu}} & = & \partial_{\mu} + ig_{\omega B}{\omega_{\mu}}
+ ig_{\phi B}{\phi_{\mu}} + ig_{\rho B}I_{3B}{\tau^a}{\rho^a_{\mu}},
\end{eqnarray}
whereas $R^a_{\mu\nu}$, and $\Omega_{\mu\nu}$ are the field tensors
\begin{equation}
R^a_{\mu\nu}= \partial_{\mu}\rho^a_{\nu} -
\partial_{\nu}\rho^a_{\mu} +
g_{\rho}\epsilon_{abc}\rho^b_{\mu}\rho^c_{\nu},
\end{equation}
\begin{equation}
{\Omega_{\mu\nu}} =  \partial_{\mu}\omega_{\nu} -
\partial_{\nu}\omega_{\mu},
\end{equation}
where $m_{B}$ denotes the baryon and $m_\sigma$, $m_\omega$,
$m_\rho$ are the masses assigned to the meson fields. Using this
Lagrangian, we derive the equation of motion and solved it in the
mean field approximation self consistently. Here,
the meson fields are replaced by their classical expectation
values. The field equations for $\sigma$, $\omega$ and $\rho$-meson
are given by
\begin{widetext}
\begin{eqnarray}
m^2_\sigma\left(\sigma_0 + \frac{g_{\sigma
B}\kappa_{3}}{2m_{B}}{\sigma^2_0} +\frac{g^2_{\sigma
B}\kappa_{4}}{6m_{B}^2}{\sigma^3_0}\right) -\frac{1}{2}m^2_{\omega}
\left(\eta_1\frac{g_{\sigma B}}{m_{B}} + \eta_2\frac{g^2_{\sigma
B}}{m_{B}^2}\sigma_0\right) {\omega^2}
-\frac{1}{2}m^2_{\rho}\eta_{\rho}\frac{g_{\sigma}}{m_{B}}\rho^2_0 =
\sum_{B}g_{\sigma B}{m^*}^2_{B}\rho_{SB}
\end{eqnarray}
\begin{eqnarray}
& & m^2_{\omega}\left(1 + \frac{{\eta_1}g_{\sigma}}{m_{B}}\sigma_0 +
\frac{{\eta_2}g^2_{\sigma}}{2m_{B}^2}\sigma^2_0\right){\omega}_0 +
\frac{1}{6}{\zeta_0} g^2_{\omega B}{\omega^3_0} = \sum_{B}g_{\omega
B}\rho_B
\end{eqnarray}
\end{widetext}
\begin{eqnarray}
& & m^2_{\rho}\left(1 + \frac{g_{\sigma
B}{\eta_{\rho}}}{m_{B}}\sigma_0\right)\rho_{03} = \sum_{B}g_{\rho
B}I_{3B}\rho_B
\end{eqnarray}
For a baryon species, the scalar density, $\rho_{SB}$, and baryon
density $(\rho_B)$ are
\begin{equation}
\rho_{SB}= \frac{2J_{B}+1}{2\pi^2}\int_{0}^{k_B} \frac{k^2
dk}{E^*_B}
\end{equation}
\begin{equation}
\rho_{B} = \frac{2J_{B}+1}{2\pi^2}\int_{0}^{k_B} {k^2 dk}
\end{equation}
where $E^{*}_{B}=\sqrt{k^{2}+{m^*}^2_{B}}$ is the effective energy
and $J_{B}$ and $I_{3B}$ are the spin and isospin projection of
baryon $B$, the quantity $k_B$ is the Fermi momentum for the baryon,
$m^*=m_B-g_{\sigma B}\sigma$ is the effective mass, which is solve
self-consistently from equation (5). After obtaining the
self-consistent fields, the pressure P and total energy density
$\varepsilon $ for a given baryon density are
\begin{widetext}
\begin{eqnarray}
P &=& \frac{\gamma}{3(2\pi )^{3}}\int_{0}^{k_B}
d^{3}k\frac{k^{2}}{E^{*}_{B}(k)}+\frac{1}{ 4!}\zeta _{0}g_{\omega
B}^{2}{\omega}_{0}^{4}+\frac{1}{2}\Bigg(1+\eta _{1}\frac{g_{\sigma
B} \sigma_{0}}{m_{B}}+\frac{\eta _{2}}{2}\frac{g_{\sigma
B}^{2}\sigma_{0}^{2}}{m_{B}^{2}}\Bigg)
m_{\omega B}^{2}{\omega}_{0}^{2}  \nonumber \\
&&\null -m_{\sigma
B}^{2}\sigma_{0}^{2}\Bigg(\frac{1}{2}+\frac{\kappa _{3}g_{\sigma
B}\sigma _{0}}{3!m_{B}}+\frac{\kappa _{4}g_{\sigma
B}^{2}\sigma_{0}^{2}}{4!m_{B}^{2}}\Bigg)+\frac{1}{2 }\Bigg(1+\eta
_{\rho }\frac{g_{\sigma B}\sigma_{0}}{m_{B}}\Bigg)m_{\rho
}^{2}\rho_{0}^{2} +\sum_{l}P_{l}\;,
\label{eqFN25}
\end{eqnarray}
\begin{eqnarray}
\varepsilon &=& \frac{\gamma}{(2\pi )^{3}}\int_{0}^{k_B} d^{3}k
E^{*}_{B}(k)+\frac{1}{ 8}\zeta _{0}g_{\omega
B}^{2}{\omega}_{0}^{4}+\frac{1}{2}\Bigg(1+\eta _{1}\frac{g_{\sigma
B} \sigma_{0}}{m_{B}}+\frac{\eta _{2}}{2}\frac{g_{\sigma
B}^{2}\sigma_{0}^{2}}{m_{B}^{2}}\Bigg)
m_{\omega B}^{2}{\omega}_{0}^{2}  \nonumber \\
&&\null +m_{\sigma B}^{2}\sigma_{0}^{2}\Bigg(\frac{1}{2}+
\frac{\kappa _{3}g_{\sigma B}\sigma_{0}}{3!m_{B}}+\frac{\kappa
_{4}g_{\sigma B}^{2}\sigma _{0}^{2}}{4!m_{B}^{2}}\Bigg)+\frac{1}{2
}\Bigg(1+\eta _{\rho }\frac{g_{\sigma B}\sigma
_{0}}{m_{B}}\Bigg)m_{\rho }^{2}\rho_{0}^{2}
+\sum_{l}\varepsilon_{l}\;,
\end{eqnarray}
\end{widetext}
As explained earlier, the terms in equations (10) and (11) with the
subscript $"B"$ should be interpreted as sum over the states of all
baryonic octets and $\gamma=2$ is the spin degeneracy. In the above,
$P_{l}$ and $\varepsilon_{l}$ are lepton pressure and energy density
respectively, explained in the following subsections.

For stars in which the strongly interacting particles are baryons,
the composition is determined by the requirements of charge
neutrality and $\beta$-equilibrium conditions under the weak
processes $B_1 \to B_2 + l + {\overline \nu}_l$ and $B_2 + l \to B_1
+ \nu_l$. After deleptonization, the charge neutrality condition
yields
\begin{equation}
q_{\rm tot} = \sum_B q_B (2J_B + 1) k_B^3 \big/ (6\pi^2) +
\sum_{l=e,\mu} q_l k_l^3 \big/ (3\pi^2)  = 0 ~, \label{neutral}
\end{equation}
where $q_B$ corresponds to the electric charge of baryon species $B$
and $q_l$ corresponds to the electric charge of lepton species $l$.
Since the time scale of a star is effectively infinite compared to
the weak interaction time scale, weak interaction violates
strangeness conservation. The strangeness quantum number is
therefore not conserved in a star and the net strangeness is
determined by the condition of $\beta$-equilibrium which for baryon
$B$ is then given by $\mu_B = b_B\mu_n - q_B\mu_e$, where $\mu_B$ is
the chemical potential of baryon $B$ and $b_B$ its baryon number.
Thus the chemical potential of any baryon can be obtained from the
two independent chemical potentials $\mu_n$ and $\mu_e$ of neutron
and electron respectively.

The lepton Fermi momenta are the positive real solutions of $(k_e^2
+ m_e^2)^{1/2} =  \mu_e$ and $(k_\mu^2 + m_\mu^2)^{1/2} = \mu_\mu =
\mu_e$. The equilibrium composition of the star is obtained by
solving the set of Eqs. (5)- (7) in conjunction with the charge
neutrality condition  (\ref{neutral}) at a given total baryonic
density $\rho = \sum_B (2J_B + 1) k_B^3/(6\pi^2)$; the baryon
effective masses are obtained self-consistently.

\subsection {Unpaired Quark Matter}
In the central part of neutron star the density is expected to high
enough that hadronic matter undergoes to quark degrees of freedom. In quark
phase we employ MIT bag model to describe unpaired quark matter
\cite{Ch74,Fr78,Fa84}. The bag model provides a useful
phenomenological description of quarks being confined inside the
hadrons. Confinement results from the balance of the bag pressure on
the bag walls from the outside and the pressure resulting from the
kinetic energy of the quarks inside the bag. For the quark matter we
use the EOS of ref.\cite{Fa84} in which the u,d and s quark are
degrees of freedom with electron. In this model the masses of u and
d are set to 5.0 MeV and strange quark mass is taken to be 150 MeV.
The chemical equilibrium is given by
\begin{eqnarray}
\mu_{d} =  \mu_{s} = \mu_{u} + \mu_{e}
\end{eqnarray}
The $\mu_{n}$ and $\mu_{e}$ are the two independent chemical
potentials and rest can be written in terms of them as follows:

\begin{eqnarray}
\mu_{u}=\frac{1}{3}\mu_{n}-\frac{2}{3}\mu_{e}
\end{eqnarray}
\begin{eqnarray}
\mu_{d}=\frac{1}{3}\mu_{n}+\frac{1}{3}\mu_{e}
\end{eqnarray}
\begin{eqnarray}
\mu_{s}=\frac{1}{3}\mu_{n}+\frac{1}{3}\mu_{e}
\end{eqnarray}

The pressure for quark flavor $f$, with $f$= $u$,$d$ or $s$ is \cite{Ka}
\begin{eqnarray}
P_{q} = \frac{1}{4\pi^2}\sum_{f}[\mu_{f}k_{f}(\mu^2_{f}-\frac{5}{2}m^2_{f})
+\frac{3}{2}m^4_{f}\ln{\frac{\mu_{f}+k_{f}}{m_{f}}}]
\end{eqnarray}
where $k_{f}= (\mu^2_{f}-m^2_{f})^{1/2}$ is the fermi momentum.

The leptons pressure is
\begin{eqnarray}
P_{l} & = & \frac{1}{3\pi^2}\sum_{l}\int \frac{p^4dp}{({p^2+m^2_{l}})^{1/2}}
\end{eqnarray}

The total pressure is given by
\begin{eqnarray}
P & = & P_{l} + P_{q} - B
\end{eqnarray}
where $B$ is the bag constant.

\subsection{Color Flavor Locked Quark Matter}

In this section, we consider the quark matter phase as color
flavor locked (CFL) quark paired, in which the quark
near the Fermi surface form Cooper pairs which condense,
breaking the color gauge symmetry.
In CFL phase, all the three colors and three flavors are allowed
to pair and this is the favored  phase at sufficiently high density.
We describe the CFL phase by using the thermodynamical potential

\begin{widetext}
\begin{eqnarray}
\Omega_{CFL}\left(\mu_{q},\mu_{e}\right)  =
\Omega_{quarks}\left(\mu_{q}\right) +
\Omega_{GB}\left(\mu_{q},\mu_{e}\right)+\Omega_{l}\left(\mu_{e}\right),
\end{eqnarray}
where $\mu_{q}=\frac{\mu_{n}}{3}$ and
\begin{eqnarray}
\Omega_{quarks}\left(\mu_{q}\right) =
\frac{3}{\pi^2}\sum_{f=u,d,s}\int_{0}^{\nu} p^2dp \left(\sqrt{p^2 +
m^2_{f}}-\mu_{q}\right)-\frac{3\Delta^2\mu^2_{q}}{\pi^2} + B
\end{eqnarray}
\end{widetext}
\begin{eqnarray}
\nu & = &\sqrt{\mu^2_{q}+\frac{m^2_{u}}{3}} +
\sqrt{\mu^2_{q}+\frac{m^2_{d}}{3}} -\sqrt{\mu^2_{q}+\frac{m^2_{s}}{3}}
\end{eqnarray}
and the Goldstone boson contribution due to chiral symmetry breaking
in the CFL phase is given by

\begin{eqnarray}
\Omega_{GB}\left(\mu_{q},\mu_{e}\right)  =  -\frac{1}{2}f^2_{\pi}\mu^2_{e}
\left(1-\frac{m^2_{\pi}}{\mu^2_{e}}\right)^2
\end{eqnarray}
where the paprameters are
\begin{eqnarray}
f^2_{\pi} = \frac{\left(21-8\ln 2\right)\mu^2}{36\pi^2},
\end{eqnarray}
\begin{eqnarray}
m^2_{\pi} =  \frac{3\Delta^2}{\pi^2 f^2_{\pi}}m_{s} \left(m_{u}+m_{d}\right)
\end{eqnarray}
and the electron contribution $\Omega_{l}$ is given by
\begin{eqnarray}
\Omega_{l}\left(\mu_{e}\right)  =  -\frac{\mu^4_{e}}{12\pi^2}
\end{eqnarray}

The quark number densities and the electrical charge density
carried by the pion is given by
\begin{eqnarray}
\rho_{u} = \rho_{d} = \rho_{s} = \frac{\nu^3 + 2\Delta^2\mu_{q}}{\pi^2}
\end{eqnarray}
where $\Delta$ is gap parameter and its value is 100 MeV, which is
the typical value considered in the literature.

The electric charge density carried by the pion condensate is given
by
\begin{equation}
Q_{CFL}=-f_\pi^2 \mu_e \left(1 - \frac{m_\pi^4}{\mu_e^4}
\right).\label{qcf}
\end{equation}
In the above thermodynamic potential, we have neglected the
contribution due to the kaon condensation which is an effect of
order  $m_s^4$  and thereby small compared to the$\Delta^2\mu_q^2$
contribution to the thermodynamic potential for $\Delta\sim 100$
MeV.

\subsection{Mixed Phase and Star Properties}

The mixed phase is obtained by applying charge neutrality condition
and Gibbs criteria for hadron and quark phase. The charge neutrality
condition is:
\begin{eqnarray}
 \chi\rho^{QP}_{c} + (1-\chi)\rho^{HP}_{c} + \rho^l_{c} & = & 0
\end{eqnarray}
where $\rho^{QP}_{c}$ and $\rho^{HP}_{c}$ are the charge density of
quark and hadron phase, $\chi$ and $(1-\chi)$ are the volume
fraction occupied by quark and hadron phase respectively. The phase
boundary of the coexistence region between the hadron phase and
quark phase is determined by Gibbs criteria.The critical pressure,
critical neutron and electron chemical potentials are determined by
the conditions:
\begin{eqnarray}
\mu_{HP,i} & = & \mu_{QP,i} = \mu_{i}, i=n,e
\end{eqnarray}
\begin{eqnarray}
T_{HP} =  T_{QP},
\end{eqnarray}
\begin{eqnarray}
P_{HP}(\mu_{HP},T) =  P_{QP}(\mu_{QP},T),
\end{eqnarray}
These are the chemical, thermal and mechanical equilibrium,
respectively. The energy and the total baryon densities in the mixed
phase are:
\begin{eqnarray}
\varepsilon & = & \chi\varepsilon^{QP} + (1-\chi)\varepsilon^{HP} + \varepsilon^{l},
\end{eqnarray}
and
\begin{eqnarray}
\rho & = & \chi\rho^{QP} + (1-\chi)\rho^{HP}.
\end{eqnarray}
Once the above quantities are obtained, we can construct the EOS for
mixed phase and consequently compute the properties of neutron star.
To evaluate the star structure we use the Tolman-Oppenheimer-Volkoff
(TOV) equations  found in Ref.\ \cite {To39}. They are
\begin{eqnarray}
\frac{dP}{dr} & = & -\frac{G}{r}\frac{[\epsilon + P][M + 4\pi r^3 P]}{r-2GM},
\end{eqnarray}
\begin{eqnarray}
\frac{dM}{dr} & = & 4\pi r^2\varepsilon
\end{eqnarray}
where $G$ and $M(r)$ are the gravitational constant and enclosed
gravitational mass and $c=1$. For a given EOS, these equations can
be integrated from the origin as an initial value problem for a
given choice of the central density, $(\varepsilon_0)$. The value of
$r(=R)$ at which the pressure vanishes defines the surface of the
star.

\section{RESULTS AND DISCUSSIONS}

We obtain the EOS for hadronic matter by changing incompressibility
K=300 MeV and effective mass (M*/M) to 0.7 of G2 parameter set \cite
{Es01}. It is to be noted that we are not getting any mixed phase
with original G2 set. So we change the incompressibility K from 215
MeV to 300 MeV and effective mass (M*/M)  from 0.664 to 0.7. The
resultant parameter set satisfies all the nuclear saturation
properties so that our extrapolation to higher density remains
meaningful. After the above re-adjustment, the modified Lagrangian 
parameters are given in Table I. To compare the changes, the original
G2 set is also displayed in the Table. 
\begin{widetext}
\begin{center}
\begin{table}[h]
\caption{G2 and modified G2 parameter set (G2*)}
\begin{tabular}{|c|c|c|c|c|c|c|c|c|c|c|c|c|c|c|}
\hline
Set & $m_s/M$ & $m_{\omega}/M$ & $m_{\rho}/M$ & $g_s/4{\pi}$ &
$g_{\omega}/4{\pi}$ & $g_{\rho}/4{\pi}$ & $\kappa_{3}$ & $\kappa_{4}$
& $\zeta_{0}$ & $\eta_{1}$ & $\eta_{2}$ & $\eta_{\rho}$\\
\hline
\hline
G2* & 0.554 & 0.833  & 0.820 & 0.751 &
0.936 & 0.614 & 1.202 & 14.981
& 2.642 & 0.650 & 0.110 & 0.390\\
G2 & 0.554 & 0.833 & 0.820 & 0.835 &
1.016 & 0.755 & 3.247 & 0.632
& 2.642 & 0.650 & 0.110 & 0.390\\

\hline
\end{tabular}
\end{table}
\end{center}
\end{widetext}
We assume that all the hyperons in the octet have the same
couplings. They are expressed as a ratio to the nucleon
coupling $x_{\sigma}=x_{H\sigma}/x_{N\sigma}=\sqrt\frac{2}{3}$,
$x_{\omega}=x_{H\omega}/x_{N\omega}=\sqrt\frac{2}{3}$ and
$x_{\rho}=x_{H\rho}/x_{N\rho}=\sqrt\frac{2}{3}$. 
Using the modified values of G2 set of Table I, the EOS for 
quark matter obtained with unpaired quark model and
color flavor locked phase for different values of the Bag pressure
B. In our calculation, we took  $B^{\frac{1}{4}} = 170$ MeV, 188 MeV
for UQM and CFL models respectively, and then calculate the equation
of state. For the CFL, it is to be noted that we do not find any
mixed phase at $B^{\frac{1}{4}}$ = 170 MeV and 180 MeV. In both the
cases we have $m_{u}=~m_{d} =~5$ MeV and $m_{s}=~ 150$ MeV.

In Fig. 1 we plot the resulting EOS for both E-RMF+UQM and E-RMF+CFL
cases. From Fig. 1 it is evident that the mixed phase starts earlier
in case of UQM whereas CFL predicts the mixed phase at higher
density approximately at around $\varepsilon = 2$ fm$^{-4}$. The
inclusion of hyperons softens the EOS \cite{Pr97} of the charge
neutral dense matter. This is clearly seen from the change in the
slope of Fig. 1 for energy density greater than $\varepsilon
~\approx 2.2$ fm$^{-4}$.

In Fig.2, the fractional particle densities, $\rho_i/\rho_0$ for
baryons, leptons and quarks in E-RMF+UQM are shown. From the Fig.2,
we notice the onset of quarks at around 1.3$\rho_0$ and immediately
in the vicinity of 4$\rho_0$ they are the most abundant particle in
the matter. With increasing density the model predicts pure quark
matter after 6$\rho_0$. At this density and thereafter all the
three quarks under consideration contributes almost equally to the
matter density. As evident from the figure that the leptonic
contribution ceases after 4$\rho_0$, which are primarily used up to
maintain charge neutrality of the matter. The nucleons constitutes a
sizeable population in the matter but after they decrease abruptly
at around 5$\rho_0$, after which its a pure quark phase. Further it
can be seen that $\Lambda$ appears at 4$\rho_0$ and with the
decrease of nucleons the particle fraction decreases, finally ends
at around 6$\rho_0$. The appearance of $\Lambda$ is decided purely
by the neutron chemical potential since it is isospin independent
and hence the density of $\Lambda$ decreases with decrease in
neutron density. However the contribution of $\Lambda$ is quite
negligible overall.

In Fig.3, the particle population for the baryons, leptons and
quarks are shown as a function of baryon density up to 8$\rho_0$ in
case of E-RMF+CFL for $B^{\frac{1}{4}} = 188 MeV$. From the Fig., it
can be seen that the appearance of quark starts $\sim$2.5$\rho_0$. 
It is to be noted that the quarks are of equal densities
in CFL. The deleptonization in the matter occurs at $\sim$5.5$\rho_0$. 
Similarly, the nucleons follows similar trend and at
$\sim$5.5$\rho_0$ the matter is in pure quark phase. Like Fig.2,
$\Lambda$ appears at 4$\rho_0$ and decrease with the nucleons
particle fractions and finally ends at 5.5$\rho_0$. In CFL case,
$\Sigma^{-}$ appears at 1.89 $\rho_0$ and ends at 3.5$\rho_0$. In
both the cases i.e., E-RMF+UQM and E-RMF+CFL the difference is due
to the different charge neutrality conditions.

The strangeness content in case of high density matter is an
exciting possibility, which can give important insights into some
of the most fundamental problems of astrophysics and high density behavior.
Figure 4 displays the strangeness content in the core and the crust of
the neutron star as obtained in our calculations. The strangeness content
is given by
\begin{eqnarray}
r_s & = & \chi r_s^{QP} + (1-\chi) r_s^{HP}.
\end{eqnarray}
with
\begin{eqnarray}
r_s^{QP}=\frac{\rho_s}{3\rho}, r_s^{HP}=
\frac{\sum_B|q_s^{B}|\rho_B}{3\rho}
\end{eqnarray}
where $q_s^{B}$ is the strange charge of baryon B. In both the
cases, the strangeness fraction rises steadily. However in case of
UQM the onset of strangeness starts at around 1.3 $\rho_0$ and gets
saturated at around 6.0 $\rho_0$ where the matter enters in a pure
quark phase. Similarly, the strangeness content in CFL starts at
around 2 times normal nuclear matter density, then rises steadily in
the mixed phase and ultimately gets saturated at $\sim$6.6$\rho_0$.
Using the EOS corresponding to the UQM and CFL we now compute the
properties of hybrid neutron star similar to the case of pure
neutron star mass, $M_\odot$ and
radius, gravitational red shift ($Z$), the baryonic mass of the
star, $M_b$.

In Fig. 5 the maximum mass of the hybrid star is plotted as a function of
the radius of the star as obtained by the UQM and CFL model in our
calculations. The radius of the maximum mass of the star is sensitive to
the low density EOS. In order to calculate the radius and to plot it
versus the star mass, we have used the results of Baym, Pethick and
Sutherland \cite{Ba71} for low baryonic densities. The maximum mass
obtained are 1.44 $M_\odot$ and 1.35 $M_\odot$ for UQM and CFL
respectively. The corresponding radius obtained are 9.35 km and 9.33
km respectively. These results are found to be in good agreement
with different field theoretical models and also from observational
point of view\cite{Co02,Sa02}. The success of the model is evident 
from figure 7, where we plot M/R ratios as reported by \cite{Co02,Sa02}.

Constraints on the mass-to-radius ratio can be obtained from
accurate measurements of the gravitational red shift of spectral
lines produced in neutron star photospheres. As already mentioned
that a red shift of 0.35 from three different transitions of the
spectra of the X-ray binary EXO0748-676 was obtained \cite{Co02},
which corresponds to $M/R = 0.15 M_{\odot}/km$. Another constraint
to the mass-to-radius ratio given by $M/R = 0.069M_{\odot}/km$ to
$M/R = 0.115M_{\odot}/km$ was determined from the observation of two
absorption features in the source spectrum of the 1E 1207.4-5209
neutron star \cite{Sa02}. However, in the second case, the
interpretation of the absorption features as atomic transition lines
is controversial. The absorption features are of cyclotron nature
which make the related constraints unrealistic \cite{Bi03,Xu03}. In
our calculations the M/R ratio comes out to be 0.15 and 0.14 for UQM
and CFL respectively. It is evident that our results are in very
good agreement with the observed analysis. We have added the lines
corresponding to those constraints shown by straight lines in Fig.5.

Baryonic mass as a function of maximum mass of the star is plotted
in Fig. 6. It is to be noted that the baryon mass always exceeds the
maximum mass, which is typical of the compact objects. The
difference between the two is defined as the gravitational binding
of the star. The baryonic mass, $M_b (M_\odot)$, obtained by UQM and
CFL are 1.63 $M_\odot$ and 1.51 $M_\odot$ respectively.

Gravitational red shift of a star is an important property that is
defined as
\begin{eqnarray}
Z = \frac{1}{\sqrt{1-2GM/Rc^2}}.
\end{eqnarray}
Red shift primarily depends on the M/R ratio of the star and
observationally in absence of precise and direct measurements of
mass and radius of the star it serves the purpose to constrain the
EOS of neutron stars. In Fig. 7, we plot the gravitational red shift
as a function of mass of the star with the straight lines as
observational constraints. The two solid straight lines are
corresponds to red shift $Z=0.12-0.23$ was determined from the
observation of two absorption features in the source spectrum of the
1E 1207.4-5209 neutron star \cite{Sa02} and the dotted dash line is
correspond to $Z=0.35$ which is determined from three different
transitions of the spectra of the X-ray binary EXO0748-676
\cite{Co02}. In our calculation, we obtained $Z=0.35$ and $Z=0.32$
for UQM and CFL respectively. Our results are in good agreement with
the data.
For the sake of completeness, we present the overall results in
Table II.
\begin{table}
\caption{Calculated properties of Neutron star}
\begin{tabular}{|c|c|c|c|c|c|c|}
\hline
EOS & $M (M_\odot)$ & $E_c$ ($10^{14}gcm^{-3}$) & $R$(km) & $M_b(M_\odot)$ & $Z$ \\
 \hline
 \hline
 UQM  & 1.44 & 31.0 & 9.35 & 1.62 & 0.35 \\
 CFL  & 1.35 & 32.0 & 9.33 & 1.51 & 0.32 \\
 \hline
\end{tabular}
\end{table}

\section{SUMMARY AND CONCLUSIONS}
In the summary, we obtained the EOS of state for neutron star
and studied the various properties of the star i.e., like mass, radius and
redshift. The phase transitions
from hadron to quark matter and the existence of the mixed phase
in the inner core of the star are analyzed. The considered
model based on the assumption that the neutron star consists
of the inner core and crusts as two main parts of it. The 
theoretical model \cite{Fu96} for the outer region is described by
the Lagrangian which includes the full octet of baryons
interacting through the exchange of meson fields. The chosen
parameter set based on effective field theory motivated 
relativistic mean field includes
nonlinear scalar-vector and vector-vector interaction terms
which leads to the soft EOS. This is in agreement of DBHF
results. But the mixed phase EOS are not feasible, 
within the original G2 parameter set of the E-RMF formalism. 
To get a reasonable EOS for the mixed phase, for the
hadronic matter, we changed the incompressibility marginally.

The inner core of the star is described by UQM and 
CFL models. In UQM the quarks are treated as massless particles
inside a bag of finite dimension and confinements results
from the balance of the pressure on the bag walls from the
outside and the bag pressure resulting from the kinetic 
energy of the quarks inside the bag. The CFL phase consists
of equal numbers of u,d and s quark and so requires no
electrons to make it electrically neutral. In this
paper we used different bag pressure $B^{\frac{1}{4}}$= 170 MeV
and 188 MeV for UQM and CFL model, respectively.

The EOS are then employed to study and evaluate the global properties of
the hybrid neutron star like mass, radius and gravitational redshift $Z$.
The masses are predicted to be 1.44 and 1.35 solar mass and radii
9.35 and 9.33 km in E-RMF+UQM and E-RMF+CFL models, respectively. We
have also shown that the EOS with UQM and CFL quark phase satisfies
the constraint imposed by the recently measured redshift of 0.35
from three different transitions of the spectra of the X-ray binary
EXO0748-676 \cite{Co02}. The redshifts are 0.35 and 0.32 in the
E-RMF+UQM and E-RMF+CFL formalism, respectively, which coincides
very well with the measurement.

\section*{ACKNOWLEDGMENTS}
One of the authors (PKP) is thankful to Institute of Physics
for support of this research work. This work was partially 
supported by CSIR, Govt. of India under the project 
No. 03(1060)/06/EMR-II.

\begin{figure*}[tbp]
\includegraphics[angle=0,width=12.2cm, clip=true]{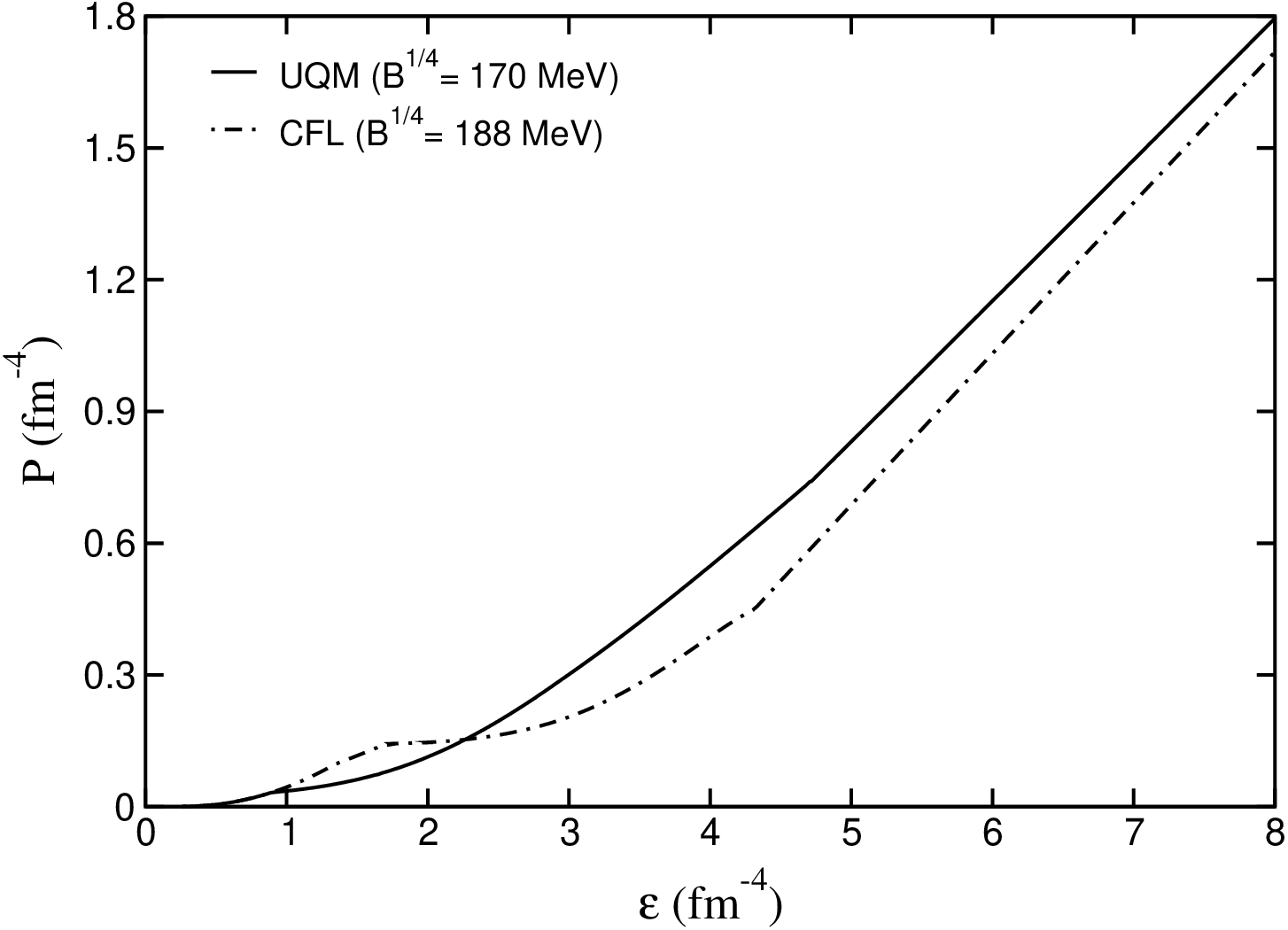}
\caption{Equation of state obtained with E-RMF model plus UQM
(solid line) and plus CFL (dash-dotted line).}
\end{figure*}

\begin{figure*}[tbp]
\includegraphics[angle=0,width=12.2cm, clip=true]{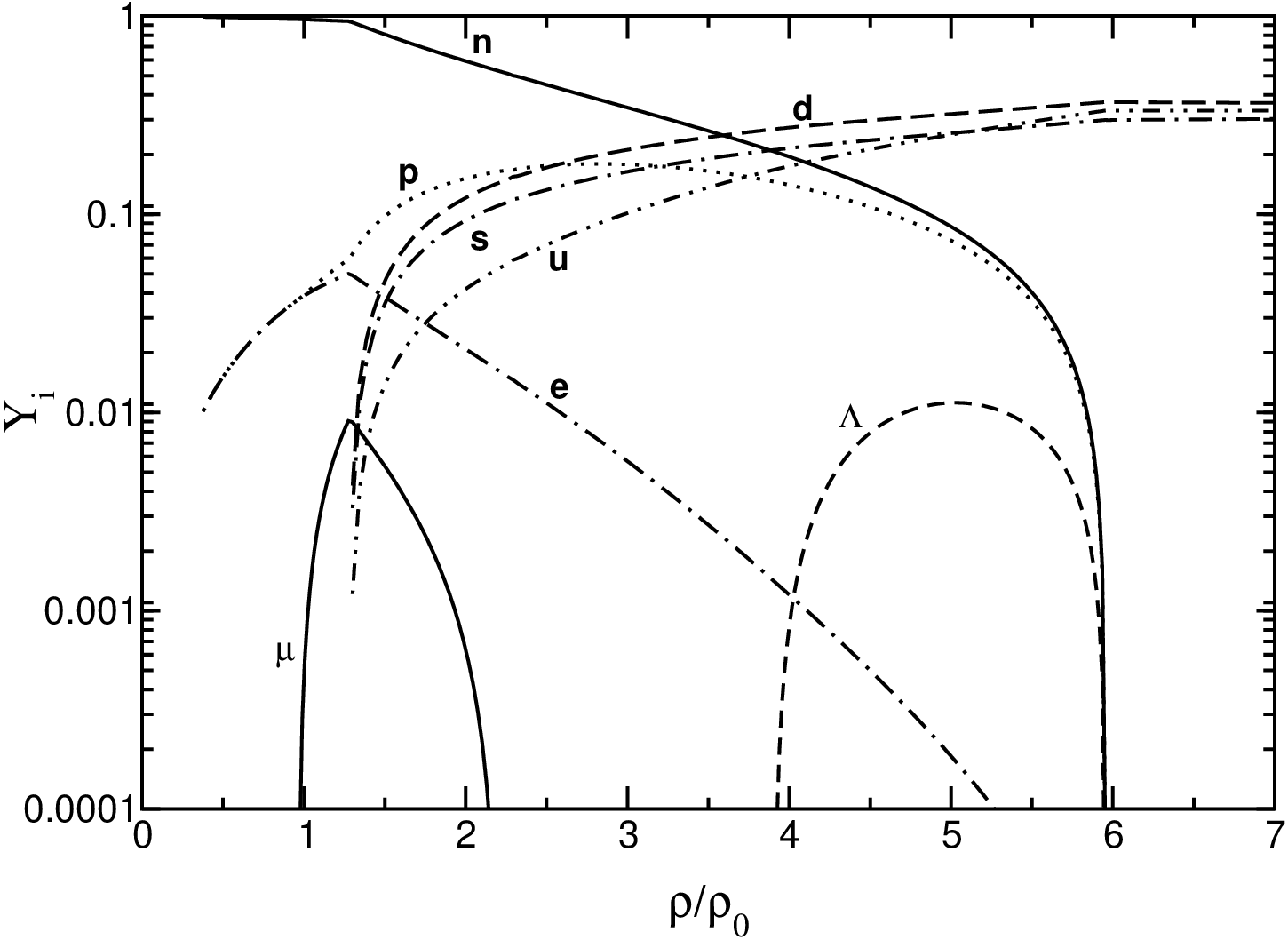}
\caption{Particle fractions, $Y_i$=$\rho$/$\rho_i$ for i = baryons,
leptons and quarks, obtained with E-RMF+UQM ($B^{\frac{1}{4}}= 170MeV$).}
\end{figure*}

\begin{figure*}[tbp]
\includegraphics[angle=0,width=12.2cm, clip=true]{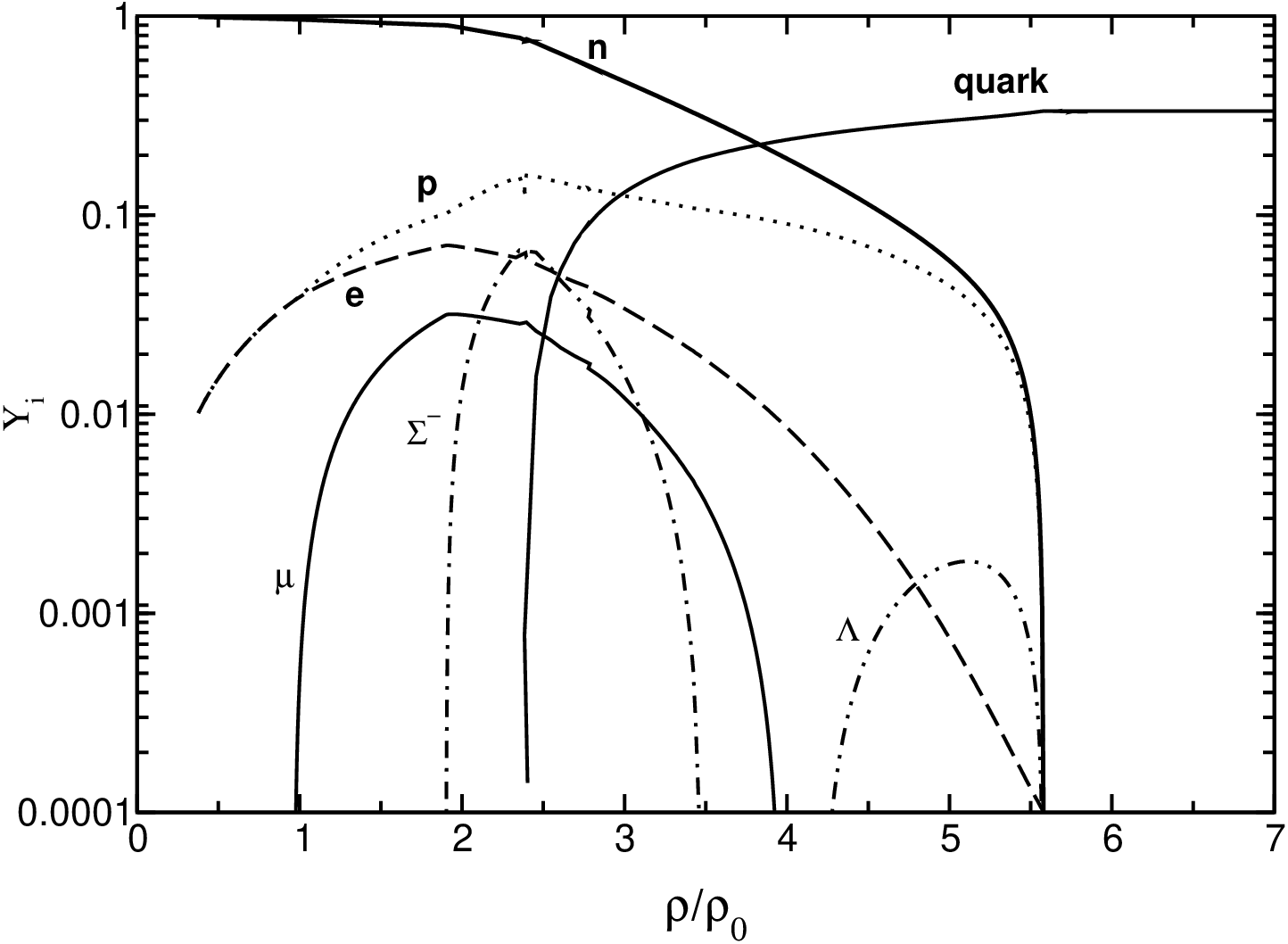}
\caption{Particle fractions, $Y_i$=$\rho$/$\rho_i$ for i = baryons,
leptons and quarks, obtained with E-RMF+CFL ($B^{\frac{1}{4}}= 188MeV$).}
\end{figure*}

\begin{figure*}[tbp]
\includegraphics[angle=0,width=12.2cm, clip=true]{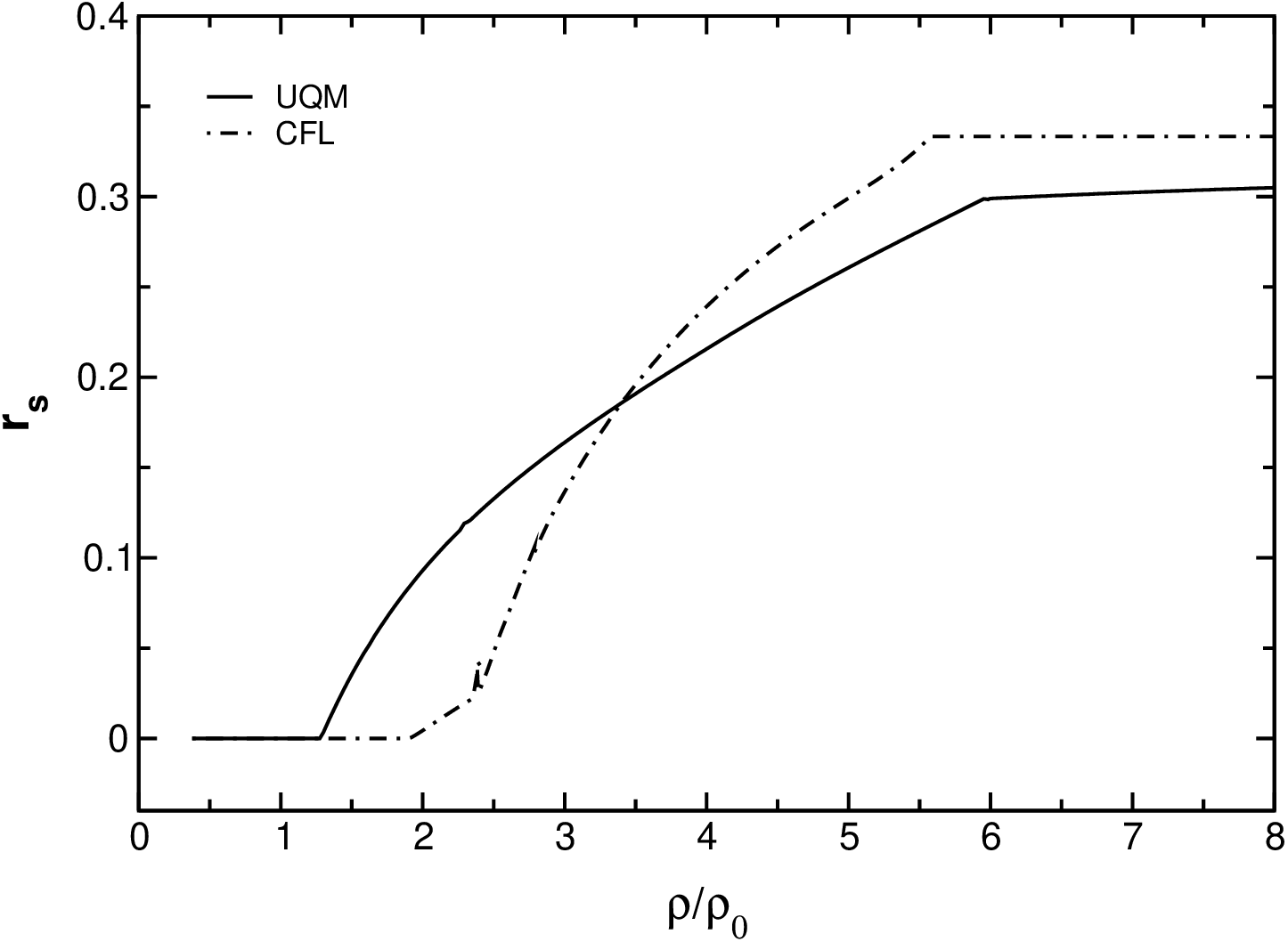}
\caption{Strangeness content obtained with E-RMF plus UQM (solid line), 
E-RMF plus CFL (dash-dotted line).}
\end{figure*}

\begin{figure*}[tbp]
\includegraphics[angle=0,width=12.2cm, clip=true]{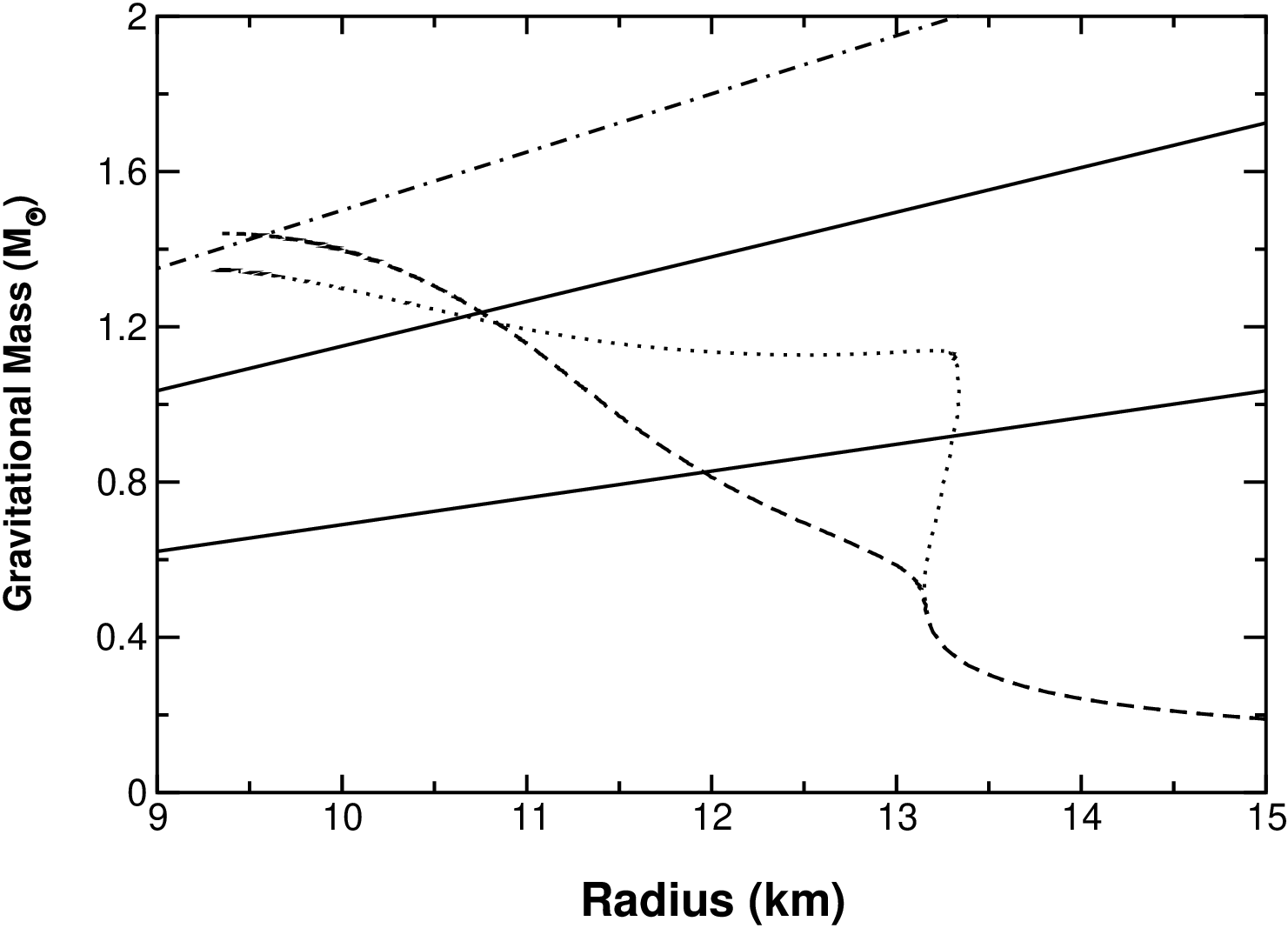}
\caption{Neutron star mass versus radius with E-RMF plus UQM (dashed
line), E-RMF plus CFL (dotted line) with experimental observations 
(solid and dash-dotted lines).}
\end{figure*}

\begin{figure*}[tbp]
\includegraphics[angle=0,width=12.2cm, clip=true]{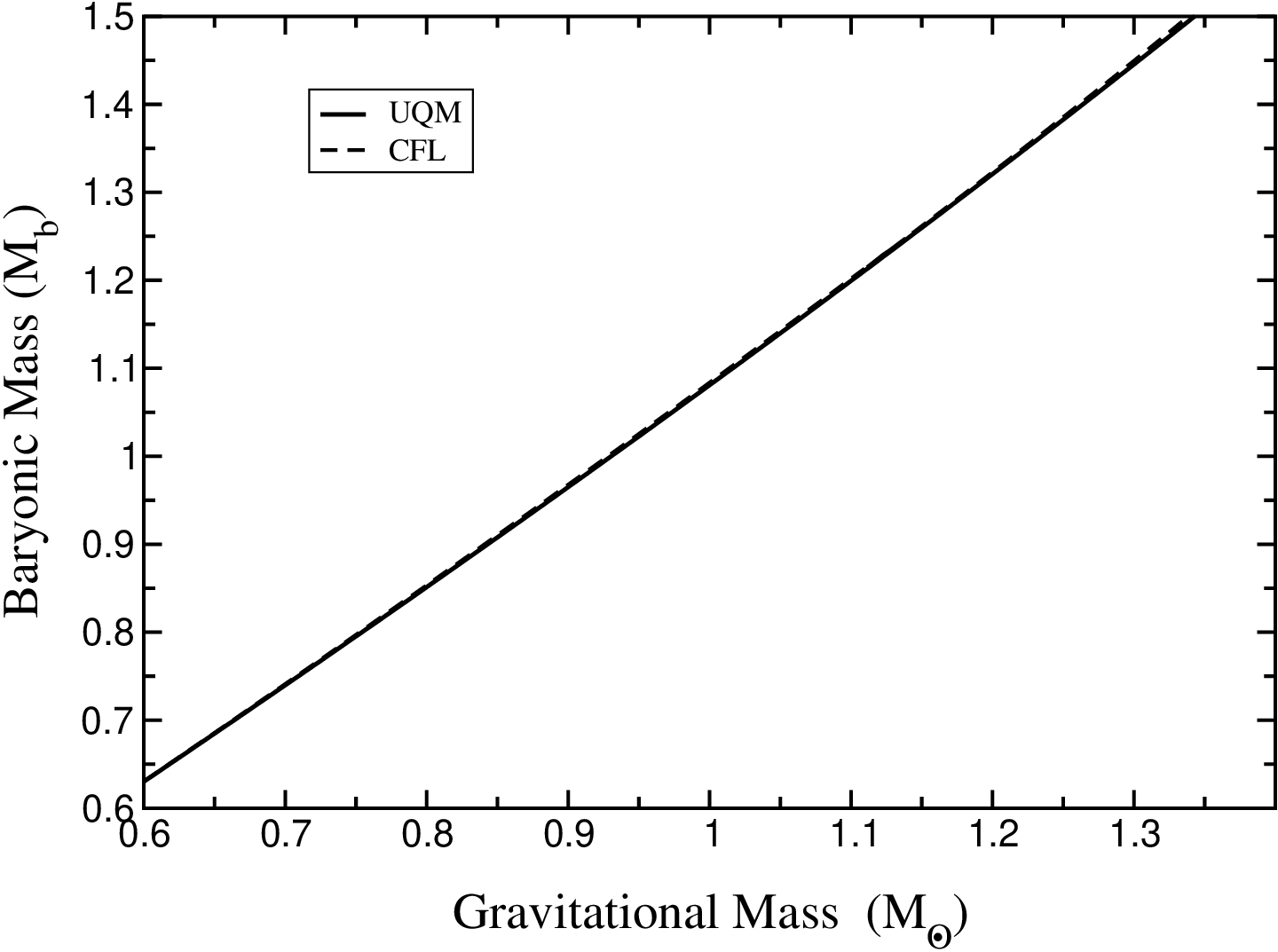}
\caption{Baryonic mass  as a function of maximum mass
of the neutron star with E-RMF plus UQM (solid line), 
E-RMF plus CFL (dashed line).}
\end{figure*}

\begin{figure*}[tbp]
\includegraphics[angle=0,width=12.2cm, clip=true]{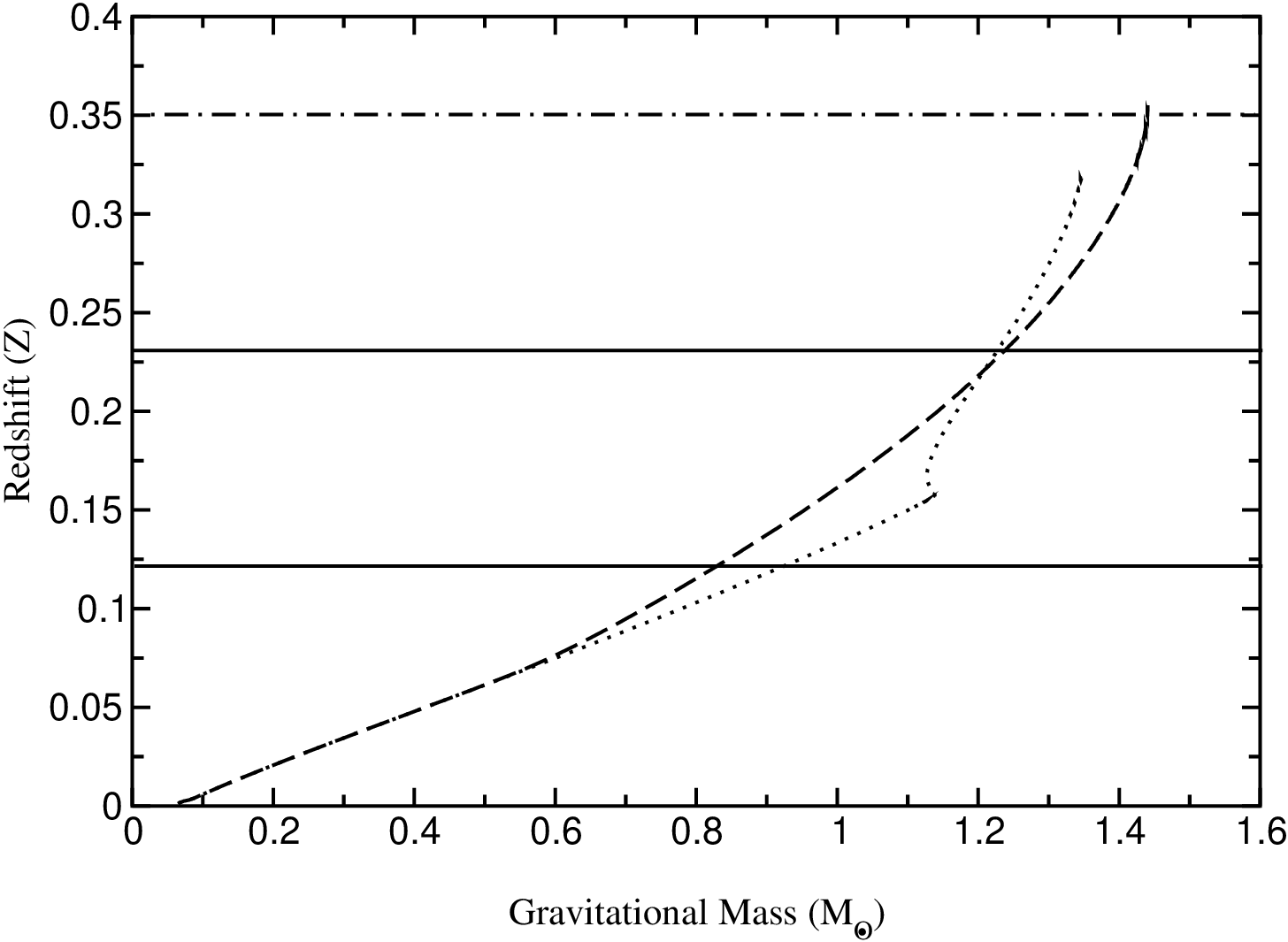}
\caption{Gravitational Redshift (Z) as a function of maximum mass
of the neutron star with E-RMF plus UQM (dashed line), E-RMF plus 
CFL (dotted line) with experimental observations
(solid and dash-dotted lines).}
\end{figure*}

\end{document}